\documentstyle[aps,prb,multicol]{revtex}

\input{psfig}

\begin{document}
\draft
\title{Theory of Electric Field-Induced Photoluminescence 
Quenching in Disordered Molecular Solids}
\author{M. C. J. M. Vissenberg$^{1,2}$ and M. J. M. de Jong$^1$}
\address{$^1$Philips Research Laboratories, 5656 AA Eindhoven, The
         Netherlands}
\address{$^2$Instituut--Lorentz, University of Leiden, 2300
         RA Leiden, The Netherlands}
\date{Phys.\ Rev.\ B {\bf 57}, 2667-2670 (1998)}

\maketitle

\begin{abstract}
The dynamics of excitons in disordered molecular solids is studied 
theoretically,
taking into account migration between different sites,
recombination, and dissociation into free charge carriers
in the presence of an electric field.
The theory is applied to interpret the results of electric field-induced 
photoluminescence (PL) quenching experiments on molecularly doped polymers by 
Deussen {\it et al.} [Chem. Phys. {\bf 207}, 147 (1996)]. 
Using an intermolecular dissociation mechanism, 
the dependence of the PL quenching on 
the electric field strength and the dopant concentration, and the 
time evolution of the transient PL quenching 
can be well described.  
The results constitute additional proof of the distinct 
exciton dissociation mechanisms in conjugated polymer blends and 
molecularly doped polymers. 
\end{abstract}

\pacs{PACS numbers: 78.55.Kz, 33.50.Hv, 71.35.Cc, 78.47.+p}

\begin{multicols}{2}

The interest in the photophysics of organic molecular solids 
is related to their broad range of  applications   
  and to the theoretical challenge of understanding 
the excited-state dynamics and transport properties 
of these disordered materials. 
The electrical and optical properties of organic dyes 
are determined by $\pi$-electrons, which 
 have a relatively small energy gap between the 
highest occupied molecular orbital (HOMO) and the 
lowest unoccupied molecular orbital (LUMO). 
The mechanical  properties in the solid state can be improved by 
dispersing the molecular dye in a polymer matrix. 
These molecularly doped polymers (MDPs) are 
broadly utilized in laser printers and
photocopiers.~\cite{Borsenberger} 
Polymers with chromophores 
built into the backbone or into side chains, 
and fully $\pi$-conjugated polymers (CPs) 
show promising potential for
applications  in thin-film devices such as
light-emitting diodes.~\cite{Burroughes}

In molecular solids, the excited states are  
strongly bound electron-hole pairs, localized on a molecular site. 
In case of energetic disorder, these excitons relax  
in the course of a random walk to sites with lower energies. 
This results in a Stokes shift between the absorption and the emission 
spectrum. 
In CPs, the $\pi$-electrons are not confined to a small molecule, 
but are delocalized over part of the polymer chain. 
Nevertheless, several spectroscopic studies   
 support  a similar molecular exciton picture for CPs.~\cite{Bassler}
Among these experiments are two complementary studies of 
electric-field induced photoluminescence (PL) quenching, 
which provide a direct comparison between 
 a CP~\cite{Kersting,Deussen} and a MDP.~\cite{Deussen2} 
The system consists of a polycarbonate matrix, blended with 
 poly(phenyl-{\it p}-phenylene vinylene) (PPPV) or with 
tris(stilbene)amine (TSA), which has a chemical structure related to that of
PPPV (see inset to Fig.~\ref{QvsT}).
In both systems, the dynamics of  the PL quenching indicates 
that the effect is due  
to the field-induced dissociation of a molecular exciton 
into separate charge carriers.~\cite{Kersting,Deussen2}  
However, clear differences have been observed in the concentration
dependence as well as the ultrafast transient behavior
of the electric field-induced PL quenching. 

In the CP blends,~\cite{Kersting,Deussen} 
(a) the PL quenching at very low PPPV concentrations is small, but finite, 
as becomes evident at high fields; 
(b) the PL quenching first increases and then saturates
with increasing concentration;
(c) the transient PL quenching keeps increasing at longer times.
Recently, we have demonstrated that all these observations 
 can be accounted for using a model that combines 
the migration of a molecular exciton   
 with an {\em intra}\/molecular dissociation mechanism, 
in which the electron and the hole are separated along the polymer 
chain.~\cite{Vissenberg} 
For instance, the finite quenching at low concentrations is a direct 
reflection of this intramolecular dissociation mechanism. 
It is also observed in transient photoconductivity
studies on CP solutions.~\cite{Gelinck}
However, our conclusion of {\em intra}\/molecular dissociation 
has been opposed by Conwell,~\cite{Conwell} who suggests an   
{\em inter}\/molecular dissociation process, 
which involves a jump of one of the 
charge carriers to a neighboring site.  
The saturation at higher concentrations  was 
ascribed to phase segregation in the polymer blend.
We have replied~\cite{VissenbergReply} that phase 
segregation can not be excluded {\it a priori}, but that the 
good agreement with experiments remains 
a strong indication of the validity of our model.

\begin{figure}
\begin{center}
\mbox{\psfig{file=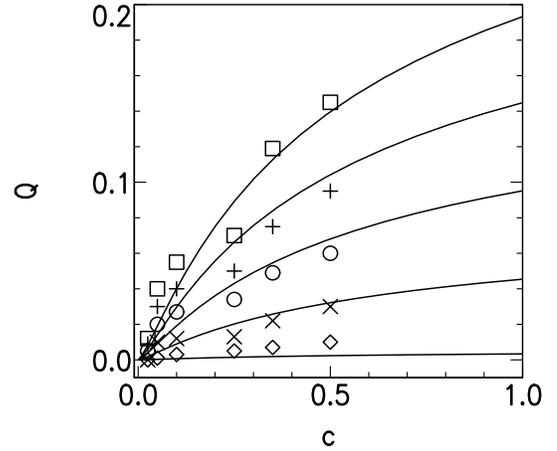,angle=0,width=7.0cm}}
\end{center}
\narrowtext
\caption{Photoluminescence quenching $Q$ as a function of concentration $c$
 for an electric field $F=0.5, 1.0, 1.5, 2.0,$  and $2.5\,$MV/cm
(bottom to top).
The parameters used are
$R=2.0\,$\AA\/ and $\delta = -7.5 \times 10^{-3}\,$eV.
Experimental data (at a temperature $T=77\,$K) 
are taken from Ref.~\protect\onlinecite{Deussen2}.}
\label{QvsC}
\end{figure}

\begin{figure}
\begin{center}
\mbox{\psfig{file=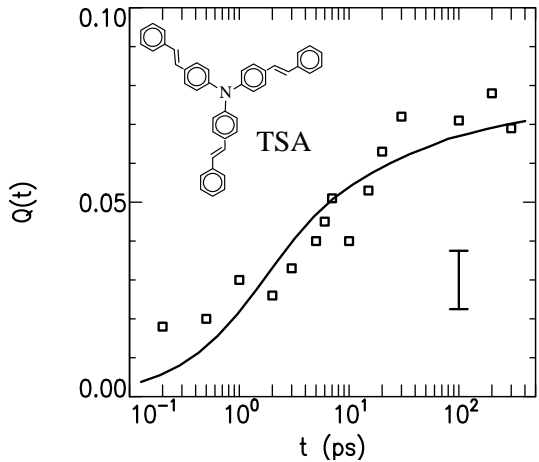,angle=0,width=7.0cm}}
\end{center}
\narrowtext
\caption{Transient photoluminescence quenching $Q(t)$ as a
function of time $t$ (solid line).
The same parameters
are used as in Fig.~\protect\ref{QvsC}, with $c = 0.5$ and $F = 1.5$ MV/cm.
Experimental data are taken from Ref.~\protect\onlinecite{Deussen2} 
($T=15\,$K). 
An indication of the experimental error is given in the lower right corner.
The chemical structure of tris(stilbene)amine (TSA) is given in the
upper left corner.}
\label{QvsT}
\end{figure}

In MDPs, where exciton dissociation must be of intermolecular 
nature, it has been observed~\cite{Deussen2} that 
(a) the PL quenching at very low TSA concentrations 
goes to zero, {\it i.e.} the probability of
dissociation on an isolated site is zero;
(b) the PL quenching increases with concentration 
over the whole range of concentrations;
(c) after an initial increase, the transient PL quenching 
saturates at longer times (see Figs.~\ref{QvsC} and~\ref{QvsT}).
In this paper, we demonstrate that  exciton migration, 
combined with {\em inter}\/molecular dissociation  of excitons, 
can account for  these observations.  
This work, combined with our previous work,~\cite{Vissenberg} 
provides a consistent picture of electric field-induced PL quenching in
 CP and MDP systems, where both the similarities and the 
differences can be explained. 

Let us first recapitulate the general theory, 
presented in Refs.~\onlinecite{Vissenberg,Movaghar,Vissenberg2}.
We consider a system consisting of localized  states $i$, with
 random positions ${\bf{R}}_i$ and  exciton energies $\varepsilon_i$,
distributed according to the density of states $\rho(\varepsilon)$.
The transition rate from site $j$ to $i$ is of the
 F\"{o}rster\cite{Forster} type, i.e.
$W_{ij} = \nu_0 {(R_0/R_{ij})}^6 \; \theta ( \varepsilon_j - \varepsilon_i)$. 
Here, the constant $R_{0}$ is the average nearest neighbor distance 
in an undiluted molecular film, 
$\nu_0$  is the corresponding nearest neighbor jump frequency, 
$R_{ij} \equiv \left| {\bf{R}}_i - {\bf{R}}_j \right|$, 
and $\theta (x)=1$ if $x>0$, $\theta (x)=0$ otherwise. 
Only jumps that are downward in energy are considered, as the 
thermal energy $k_B T$ is usually much smaller than the energy 
differences between the sites. 
The exciton decay is described by a site-independent rate $\lambda$.

Let the system be excited at $t=0$ by an incident light pulse, 
such that all sites have an initial occupational probability $f_0$. 
(Both theory~\cite{Vissenberg,Vissenberg2} and 
experiment~\cite{Deussen,Deussen2} indicate that 
the final result is insensitive to the precise initial condition.)
Then the PL intensity at time $t$ is given by
\begin{eqnarray}
L(t) & = &
 \lambda f_0 \int d\varepsilon \rho (\varepsilon) G_1 (\varepsilon, t) 
 \nonumber \\
 &&  + \lambda f_0 \int d{\bf{R}} d\varepsilon d\varepsilon' 
\rho (\varepsilon) \rho (\varepsilon')  
  G_2 ( \varepsilon, \varepsilon', {\bf R} ,t)  .
\label{eq:Lum}
\end{eqnarray}
The local Green function 
$G_1 ( \varepsilon, t)$ is
 the average probability that an exciton remains at its  
initial site,
\begin{equation}
G_1 ( \varepsilon,t)
= \exp{\left[ -\lambda t  - n(\varepsilon) {\case43}
    \pi R_0^3  \sqrt{ \pi \nu_0 t} \, \right]} ,
\label{eq:Gloc}
\end{equation}
with $n(\varepsilon) = \int_{-\infty}^{\varepsilon} d \varepsilon'
\rho ( \varepsilon' )$ 
 the density of sites with energies below $\varepsilon$. 
We note that this probability decays exponentially due to 
exciton decay, but non-exponentially due to the migration of excitons 
to different sites. 
In Eq.~(\ref{eq:Lum}), the non-local Green function 
 $G_2 ( \varepsilon,\varepsilon', {\bf{R}} ,t)$ is 
the average probability that an exciton has migrated at time $t$ to  
 a site with energy  $\varepsilon$ 
at a relative position ${\bf{R}}$ from the initial site 
with energy $\varepsilon'$. 
This non-local Green function can be decomposed into 
local Green functions, describing the different paths an exciton  
may take to end up at the final site. 
The expression for the non-local Green function is given in 
Refs.~\onlinecite{Vissenberg,Vissenberg2}. 

In the presence of an electric field ${\bf F}$, excitons may dissociate into
separate charge carriers. 
The dissociation of excitons is taken into account through 
an additional decay rate $\lambda_d$: 
on a fraction $\alpha$ of the sites, dissociation can take place 
and the excitons decay at a total rate $\lambda + \lambda_d$; 
on the remaining fraction $1-\alpha$ of the sites, 
the decay rate is just $\lambda$.
The {\em quenched} local Green function is then given by 
\begin{equation}
G_1^q (\varepsilon,t) = (1-\alpha) G_1
(\varepsilon,t) + \alpha G_1^d (\varepsilon,t) ,
\label{eq:qG}
\end{equation}
where $G_1^d (\varepsilon,t)$, {\it i.e.} the local Green function 
 on a site where dissociation can take place, 
 is given by Eq.~(\ref{eq:Gloc}) with $\lambda$ 
replaced by $\lambda + \lambda_d$. 
Substitution into Eq.~(\ref{eq:Lum}) yields 
 the time-integrated PL quenching, 
\begin{equation}
Q = \frac{\int_0^{\infty} dt \left[ L(t) -
    L^q(t) \right]}{\int_0^{\infty} dt L(t)} ,
\label{eq:quenching}
\end{equation}
as well as the transient PL quenching
\begin{equation}
Q(t) = \frac{L(t) - L^q(t)}{L(t)} ,
\label{eq:transientQ}
\end{equation}
where $L^q(t)$ denotes the quenched PL intensity in the presence of an 
electric field.

In the following, we
will assume that the process of exciton migration
is the same in both the CP blend and the MDP.
Hence we use the same  parameters
 as in Ref.~\onlinecite{Vissenberg}:
$\nu_0 =  10^{13}$ Hz,  $\lambda  =  {( 300 \; {\rm ps})}^{-1}$.
The relative concentration $c = n \case43 \pi R_0^3 \in [0,1]$ 
is proportional to the 
density of sites $n=n(\infty)$. 
The density of states $\rho (\varepsilon)$ is a Gaussian with
 standard deviation $\sigma=0.1$ eV. 
The results are not heavily dependent on the precise 
values of the parameters.

\begin{figure}[t]
\begin{center}
\mbox{\psfig{file=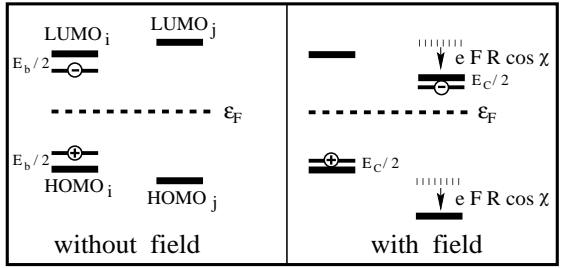,angle=-90,width=7.4cm}}
\end{center}
\narrowtext
\caption{Energy level diagram describing field-induced exciton
dissociation between two molecules $i$ and $j$.
Left figure: energy levels
of molecules $i$ and $j$ without applied electric field.
An exciton, located on molecule $i$, has an energy
$\varepsilon_i = {\rm HOMO}_i + {\rm LUMO}_i - E_b$.
Right figure:
 energy levels of  molecules $i$ and $j$ are shifted by $eFR\cos{\chi}$
in the presence of an applied electric field ${\bf F}$.
The energy of the electron-hole pair
is given by ${\rm HOMO}_i + {\rm LUMO}_j - E_C - e F R \cos{\chi}$.
(For $\cos{\chi} < 0$, the hole jumps to $j$.)
When the distribution of HOMO and LUMO levels is symmetric
with respect to the Fermi energy $\varepsilon_F$,
the energy condition~(\protect\ref{eq:energycondition}) is obtained.}
\label{energylevels}
\end{figure}

Regarding the dissociation of excitons, 
 a microscopic model of intermolecular dissociation 
 is used in order to determine the dependence of the dissociation 
parameters $\lambda_d$ and $\alpha$ on the electric field strength $F$, 
the TSA concentration $c$, and the exciton energy $\varepsilon$.
The dissociation rate for an exciton located on site $i$ with a neighbor $j$ 
is given by
\begin{equation}
\lambda_d = \nu_0 \exp{(-2 \gamma R_{ij})} \theta(E_{ij}) , 
\end{equation}
where we follow Scheidler {\it et al.}~\cite{Scheidler} in  taking 
 the attempt frequency $\nu_0=10^{13}$ Hz identical to the 
 nearest-neighbor exciton transition rate.  
The localization length is of order $\gamma^{-1}=1.5$ \AA, 
as obtained from carrier transport measurements in MDPs.~\cite{Borsenberger2}
The energy difference $E_{ij}$ 
 between an electron and a hole on separate sites 
 and an exciton on the site with lowest energy  
is given by (see Fig.~\ref{energylevels})
\begin{equation}
E_{ij} = E_{C} - E_b - 
\case12 \left| \varepsilon_i - \varepsilon_j \right| + 
e F R_{ij}  \left| \cos{\chi} \right| ,
\label{eq:energycondition}
\end{equation}
where $E_C$ is the Coulomb energy of the dissociated electron-hole pair 
and $\chi$ is the angle between the electric field ${\bf F}$ and 
the vector ${\bf R}_j - {\bf R}_i$. 
Here, we have assumed  that the exciton binding energy  
$E_b$ is the same on all sites, 
 {\it i.e.} the disorder in exciton energies  
is solely due to the inhomogeneous broadening of the  
HOMO and  LUMO levels. 
We have further assumed  a symmetric distribution of   
  HOMO and LUMO levels. 
Then a hole tunneling to a site $j$ in the field direction 
gives the same result as an electron tunneling to the same 
site $j$ in the opposite direction.

In contrast with exciton migration,  
the dissociation of excitons is a short-range process. 
We can therefore neglect the dissociation between  
sites at a distance $>R_0$, 
as these dissociation rates are 
exponentially smaller than the rates between nearest neighbors.
The probability of finding
$k$ sites within the ``nearest neighbor volume'' 
$\case43 \pi R_0^3$ around the initial site is given
by a Poisson distribution.
Hence the probability of having at least one neighbor 
 at a distance shorter than $R_0$
is $1- \exp{\left( -n\case43 \pi R_0^3 \right)}=1- e^{-c}$.
As the TSA molecule is anisotropic, 
the dissociation rate is not determined by the 
average center-to-center distance $R_0$, but by 
the minimum tunneling distance $R$, which may be much smaller.
Therefore, we replace the distribution of
rates $\lambda_d(R_{ij})$ by the dominant rate $\lambda_d(R)$.

Let us now calculate the 
probability that $E_{ij}>0$, given that 
the site $i$ has a nearest neighbor $j$.
This probability, which is a function of the site energy $\varepsilon$, 
is found by integrating over all energies $\varepsilon'$ and 
orientations $\chi$ of the nearest neighbor,
\begin{eqnarray}
p(\varepsilon) & = & \frac{1}{n} 
\int d \varepsilon' \rho(\varepsilon') 
\int_{0}^{1}  d\cos{\chi}  \nonumber \\
&& \times  \theta \left( \Delta E - 
\case12 \left| \varepsilon - \varepsilon' \right| +
e F R \cos{\chi} \right)  \nonumber \\
& = & \int_{\max{(0,2\Delta E)}}^{\max{(0,2\Delta E + 2eFR)}}
d \varepsilon'
\frac{ n(\varepsilon + \varepsilon')- n(\varepsilon - \varepsilon')}{2neFR} ,
\label{eq:kansopgeschikt}
\end{eqnarray}
where 
$\Delta E=E_C-E_b$. 
If $\Delta E<0$, all excitons are stable in the absence of an electric 
field. The stability of excitons is {\em enhanced} by the disorder in 
the system, as the excitons migrate to sites with lower energies, 
from which a dissociating jump is difficult. 
When $\Delta E+ e F R <0$, the electric field can never overcome 
the exciton binding and $p(\varepsilon)=0$.
Combining Eq.~(\ref{eq:kansopgeschikt}) with the probability to find 
a nearest neighbor, we find for the dissociation parameters 
\begin{mathletters}%
\begin{eqnarray}%
\alpha (\varepsilon) & = & 1 - e^{-c p(\varepsilon)} , \\
\lambda_d & = & \nu_0 e^{-2 \gamma R} . 
\end{eqnarray}%
\end{mathletters}%
The time-integrated PL quenching $Q$ and the transient PL quenching $Q(t)$ 
 can now be calculated as a 
function of the concentration $c$ and the electric field strength $F$,
 where the tunneling distance $R$ and the energy difference 
$\Delta E$ are the only free parameters.  

In Fig.~\ref{QvsC}, the time-integrated PL quenching $Q$ 
is shown as a function of the TSA concentration $c$ for 
different field strengths $F$.   
We find that the complete set of experiments is well described using 
 $R=2.0\,$\AA\/ and $\Delta E=-7.5 \times 10^{-3}\,$eV. 
Our theory reproduces both the zero quenching 
at low $c$ and the increasing $Q$ over the whole range of $c$. 
Deussen {\it et al.}~\cite{Deussen2} have interpreted 
their results in terms of a localization length 
$\gamma^{-1} \simeq 15-20\,$\AA. 
This value is an order of magnitude too large 
because exciton migration is not taken into account. 

The obtained values for $R$ and $\Delta E$  
 allow us to make a rough  estimate of the exciton binding energy $E_b$.
We use the approach by Scheidler {\it et al.}~\cite{Scheidler}  
\begin{equation}
\Delta E= E_C - E_b = 
\frac{e^2}{4 \pi \epsilon_r \epsilon_0 } \left( \frac{1}{R_{{\rm eh}}} - 
\frac{1}{R_{{\rm exc}}} \right) , 
\label{eq:energieverschil}
\end{equation}
with $R_{{\rm eh}}$ the electron-hole distance and $R_{{\rm exc}}$
the distance 
 between the electron and the hole in the excitonic state.
Due to the anisotropic shape of the TSA molecules, 
 the shortest dissociating jump is perpendicular to the molecule, 
$R_{{\rm eh}} = \sqrt{ R_{{\rm exc}}^2 + R^2}$. 
Substituting $\epsilon_r=3$, $R=2.0\,$\AA\/ and 
$\delta = -7.5 \times 10^{-3}\,$eV, we find 
that the exciton size $R_{{\rm exc}}$
 is about $11\,$\AA, which corresponds with
the size of one of the three ``legs'' of the TSA molecule
(see inset to Fig~\ref{QvsT}). 
Furthermore, we find $E_b=0.45\,$eV, 
which agrees well with with the value of $0.4\,$eV 
estimated in Refs.~\onlinecite{Deussen,Deussen2}.

Equation~(\ref{eq:energieverschil}) can also be used to explain the 
experimentally observed  increase in 
 $Q$ with increasing polarity of the polymer matrix.~\cite{Deussen2}
An increase in polarity  enhances    
 the relative dielectric constant $\epsilon_r$.  
Consequently, $\left|\Delta E \right|$ decreases 
and the exciton becomes more liable to dissociation. 

Let us now compare our theory with the experimental results 
on {\em transient} electric field-induced PL quenching $Q(t)$,~\cite{Deussen2} 
using the same parameters as above. 
In Fig.~\ref{QvsT}, the time evolution of  
 transient electric field-induced PL quenching $Q(t)$  is shown for 
 a 50\% TSA blend with an applied electric field $F = 1.5\,$MV/cm. 
Given the fact that we have not used any fitting parameters, the 
agreement is quite good.
We see that, after the onset of PL quenching  in the first few picoseconds, 
  $Q(t)$ first increases and then saturates at longer timescales. 
The saturation of $Q(t)$ is due to the migration of excitons 
to sites with a low energy $\varepsilon$, 
which have a very low dissociation probability $\alpha(\varepsilon)$. 
This saturation effect may be part of the answer 
to the question why
on-chain dissociation is dominant in CPs.~\cite{Conwell} 
At low concentrations, interchain dissociation is suppressed simply  
due to the large distances between the chains. 
At higher concentrations the chains are closer together, but then  
the excitons migrate rapidly to the sites with the lowest energies, 
where they are stable against intermolecular dissociation. 

In summary, we have combined our analytic theory of exciton 
migration with a microscopic model of intermolecular 
exciton dissociation, 
in order to describe electric field-induced PL quenching 
experiments on MDPs.  
The theory accounts well for the dependence of  PL quenching 
on the electric field strength and on the dopant concentration. 
We estimate an exciton binding energy $E_b=0.45\,$eV and 
an exciton size $R_{\rm exc}=11\,$\AA.
Using the same set of parameters, good agreement is obtained with 
the observed time evolution of transient PL quenching. 
This work, together with our previous work,~\cite{Vissenberg} 
provides a consistent picture of electric field-induced PL quenching 
in MDPs as well as CP blends, underlining their  
 distinct dissociation mechanisms, {\it i.e.} {\em inter}\/molecular 
dissociation in MDPs and {\em intra}\/molecular dissociation in CP blends. 

We acknowledge useful discussions with G. H. L. Brocks. 
This work has been supported by the Dutch Science Foundation NWO/FOM.

\end{multicols}

\end{document}